# Melting curve of magnesium up to 460 GPa from *ab initio* molecular dynamics simulations

Chengfeng Cui,[1,2] Jiawei Xian,[2,a] Haifeng Liu,[2] Fuyang Tian,[1,a] Xingyu Gao,[2,a] and Haifeng Song[2]

[1]Institute for Applied Physics, University of Science and Technology Beijing, Beijing, 100083, China

[2]Laboratory of Computational Physics, Institute of Applied Physics and Computational Mathematics, Beijing, 100088, China

[a]Authors to whom correspondence should be addressed:
xian_jiawei@iapcm.ac.cn, fuyang@ustb.edu.cn, gao_xingyu@iapcm.ac.cn

## ABSTRACT

Based on *ab initio* molecular dynamics simulations, we determined the melting curve of magnesium (Mg) up to ~460 GPa using the solid-liquid coexistence method. Between ~30 and 100 GPa, our melting curve is noticeably lower than those from static experiments, but is in good agreement with recent shock experiments. Up to ~450 GPa, our melting curve is generally consistent with the melting points from first-principles calculations using the small-cell coexistence method. We found that, at high pressures of a few hundred GPa, due to the strong softening of interatomic interactions in the liquid phase, solid-liquid coexistence simulations of Mg show some characteristics distinctively different from other metal systems, such as aluminum. For example, at a given volume, the pressure and temperature range for maintaining a stable solid-liquid coexistence state can be very small. The strong softening in the liquid phase also causes the unusual behavior of reentrant melting to occur at very high pressures. The onset of reentrant melting is predicted at ~305 GPa, close to that at ~300 GPa from the small-cell coexistence method. We show that the calculated melting points, considering reentrant melting, can be excellently fitted to a low-order Kechin equation, thereby making it possible for us to obtain a first-principles melting curve of Mg at pressures above 50 GPa for the first time. Similar characteristics in solid-liquid coexistence simulations, as well as reentrant melting, are also expected for other systems with strong softening in the liquid phase at high pressures.

## I. INTRODUCTION

Melting properties are of great significance in the fields of condensed matter physics, nuclear physics, Earth science, astrophysics, etc.[1-3] They are important for understanding the equilibrium properties of both solid and liquid phases of materials, as well as for establishing the multiphase equation of state (EOS).[4-7] As an important alkaline-earth metal element, magnesium (Mg) has received considerable attention for the investigation on its phase diagram,[6, 8-10] including its melting curve.[11-16]

Mg adopts the hexagonal close-packed (hcp) structure at ambient conditions.[9] At room temperature, experiments found that Mg transforms into the body-centered cubic (bcc) structure at ~50 GPa,[9, 13, 17] and the bcc phase is stable at least to 211 GPa.[13] The hcp-bcc phase line as a function of temperature was investigated in several first-principles studies using the quasi-harmonic approximation (QHA)[6, 8, 10] and measured in static experiments by Stinton *et al.*[13] At high temperatures, the transition pressures from static experiments[13] are significantly higher than all QHA results,[6, 8, 10] which may mostly originate from the neglection of lattice anharmonicity in theoretical calculations.[18, 19] Experiments also reported some evidences for the existence of the double-hexagonal-close-packed (dhcp) phase but only in a very narrow pressure range from ~10 to 18 GPa at high temperatures prior to melting.[9, 13] Several first-principles studies predicted that, at very high pressures, the bcc structure transforms into the face-centered cubic (fcc) structure, but the calculated transition pressures show large discrepancies, varying from 180 to 790 GPa at zero temperature.[10, 20-22]

At pressures below ~10 GPa, static experiments on the melting of Mg are in reasonable agreement with each other.[11, 12, 15] Using diamond-anvil-cells (DACs), Errandonea *et al*.[11] and Stinton *et al*.[13] extended the melting curve of Mg up to 80 and 105 GPa, respectively. The melting curve from Stinton *et al*.[13] overlaps with that from Errandonea *et al*.[11] at relatively low pressures, but becomes increasingly higher as the pressure rises above ~50 GPa. In particular, the melting temperature at 105 GPa from Stinton *et al*.[13] (~4400 K) is higher than the extrapolated value from Errandonea *et al*.[11] (~3600 K) by as much as ~800 K. In recent shock-wave experiments, Beason *et al*. reported two melting points at $55 \pm 2$ GPa[23] and $55.5 \pm 0.3$ GPa[16], respectively. Based on different EOS models, they estimated the corresponding melting temperatures as in the range of 2550-3120 K[23] and 2960 K,[16] respectively, both lower than the results at similar pressures from Errandonea *et al*.[11] (~3200 K) and Stinton *et al*.[13] (~3300 K).

Only two studies have investigated the melting of Mg based on first-principles simulations. Moriarty and Althoff[8] conducted free energy calculations to determine the melting curve up to 50 GPa, and their results are overall consistent with those of static experiments.[11-13, 15] Hong and van de Walle[14] calculated the melting points of the bcc phase from 60 to 450 GPa using the small-cell coexistence method, a statistical approach they developed on top of molecular dynamics (MD), but did not provide the corresponding melting curve. Their melting points are systematically lower than the melting curves from the static experiments by Errandonea *et al*.[11] and Stinton *et al*.[13] at pressures below ~100 GPa, by around 300 K and ~600-1000 K, respectively, but show good consistency with the shock melting point from Beason *et al*.[16] at close to 60 GPa, within ~100 K. Moreover, they predicted that, at high pressures above ~300 GPa, Mg exhibits the unusual behavior of reentrant melting, i.e., the melting temperature does not increase with pressure indefinitely as in the normal case, but decreases as pressure further rises.

As introduced above, at pressures between ~50 and 100 GPa, noticeable discrepancies exist in the melting of Mg, and at higher pressures there was only one first-principles study on the melting points, with no melting curve reported. In order to help resolve the above controversies and to obtain a melting curve at pressures above ~100 GPa as well, we performed *ab initio* molecular dynamics (AIMD) simulations to calculate the melting curve of the bcc phase of Mg up to ~460 GPa, using the solid-liquid coexistence method.[24] This method is one of the most reliable approaches for melting calculations,[25] in which the melting point can be determined at high accuracy through direct simulation of a stable solid-liquid coexistence state. It was demonstrated that this method can deliver results at the same level of accuracy as free energy calculations based on the formally exact thermodynamic integration (TI) method.[26-29]

At pressures between ~30 and 100 GPa, our melting curve is noticeably lower than those from the static experiments by Errandonea *et al.*[11] and Stinton *et al.*[13] but shows good agreement with the results from recent shock experiments by Beason *et al.*[23] Up to ~450 GPa, our melting curve is generally consistent with the melting points from first-principles calculations by Hong and van de Walle.[14] We find that, at high pressures of a few hundred GPa, owing to the stronger softening of interatomic interactions in the liquid than in the solid, solid-liquid coexistence simulations of Mg show some characteristics distinctively different from other metal systems, such as aluminum (Al). For example, at a given volume the pressure and temperature range for maintaining stable solid-liquid coexistence state can be very small. The strong softening in the liquid phase is also the cause for the usual behavior of reentrant melting at very high pressures. We predict that the onset of this behavior is at ~305 GPa, close to that at ~300 GPa from first-principles calculations by Hong and van de Walle.[14] We show that the melting temperature of Mg as a function of pressure, considering reentrant melting at very high pressures, can be excellently fitted to a low-order Kechin equation.[30] This allows us to obtain a first-principles melting curve of Mg at pressures above 50 GPa for the first time.

The rest of this paper is organized as follows: Section II introduces the simulation methods; Section III presents our calculated results for the melting curve of Mg and discusses the associated characteristics exhibited in solid-liquid coexistence simulations for the melting of Mg; Section IV concludes this paper.

## II. METHODOLOGY

AIMD simulations were performed via the Vienna *ab initio* simulation package (VASP),[31-33] based on the density-functional theory (DFT)[34] implemented using the projector augmented-wave (PAW) method.[35, 36] The exchange-correlation (XC) interaction was described with the generalized-gradient approximation (GGA), in the form parameterized by Perdew, Burke, and Ernzerhof (PBE).[37] A pseudopotential (Mg_pv) containing eight valence electrons [$3s^2 3p^6$] was chosen. For the high pressures considered in the present work (up to ~460 GPa), this potential has been validated by Hong and van de Walle[14] by comparing with the results calculated using the full-potential linearized augmented plane-wave (LAPW) method.[38, 39] The kinetic energy cutoff for the plane-wave expansion was set to 600 eV, ~1.5 times the default value for the chosen potential (403.929 eV). Electronic thermal excitations were accounted for using the Mermin functional formalism.[40]

Melting simulations were conducted using the solid-liquid coexistence method under the NVE (constant number of atoms, constant volume, and constant internal energy) ensemble, with a time step of 2 fs. Supercells with 640 ($4 \times 4 \times 20$) atoms were used for the bcc phase of Mg, and due to the large supercell sizes the Brillouin zone was sampled with the $\Gamma$ point only. We employed very long supercells in one direction, since this allows more freedom for the movement of the solid-liquid interface, and thus helps to maintain the solid-liquid coexistence state in the simulations.[41, 42] The threshold for the total energy convergence is $1.6 \times 10^{-8}$ eV/atom. With these prescriptions, the average energy drift in different coexistence simulations is ~2 meV/atom/ps. By comparing with simulations that have slower energy drift, achieved by reducing the time step and the total energy convergence threshold at the same time, we have confirmed that the level of energy drift in the present study has little influence on the accuracy of the calculated melting points. Coexistence simulations were performed at a series of atomic volumes ($V$) to obtain the melting temperature ($T_m$) as a function of pressure, up to ~460 GPa for the bcc phase of Mg. At each considered volume, before the much more expensive coexistence simulations, we first used the Z method[43, 44] for a quick estimation of $T_m$, as Zhang et al.[45] did in their coexistence simulations for the melting of vanadium. Through partial suppression of superheating effects, the typical overestimation of $T_m$ is only around 10% by the Z method,[42, 46] when small supercells with ~100 atoms are used, vs around 20% by the heat until it melts (HUM) method.[42] The estimated $T_m$ can serve to delimit a small range of thermodynamic conditions for locating the accurate melting point, thus saving the computational cost greatly by reducing the number of trial coexistence simulations.

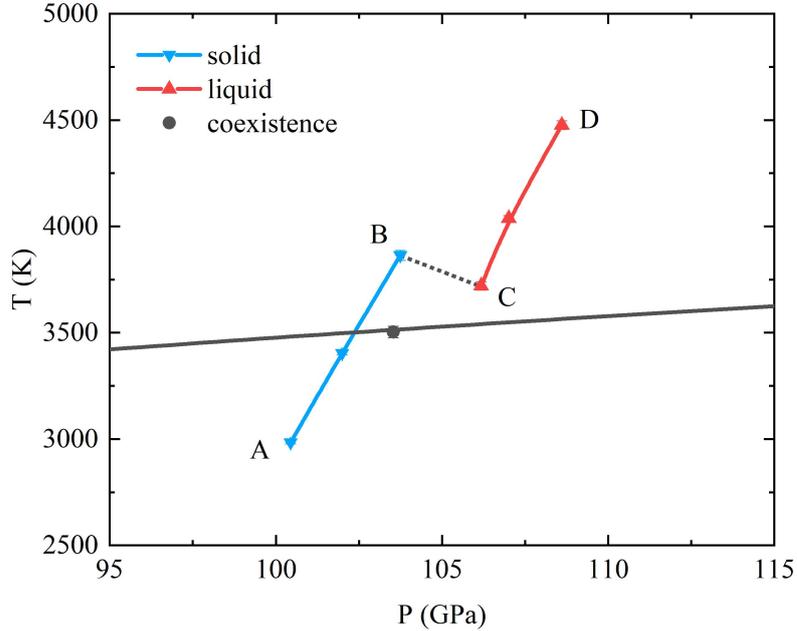

FIG. 1. Melting point estimation using the Z method, at $V = 11.6$ Å$^3$/atom for the bcc phase of Mg. The line segments "AB" and "CD" represent the isochores of the bcc and liquid phases, respectively, and the point "C" denotes the melting point estimated by the Z method. The black line is the melting curve obtained in the present study, with the black circle representing the corresponding melting point calculated using the solid-liquid coexistence method.

Figure 1 shows a typical example for the Z method, where the bcc phase of Mg at $V = 11.6$ Å$^3$/atom is considered, in a supercell of 128 ($4 \times 4 \times 4$) atoms. Several simulations were performed under the NVE ensemble, starting from the same perfect lattice configuration, but with progressively higher initial temperatures ($T_0$), corresponding to increasing internal energies ($E$). When $E$ is relatively low, the system will end in a state on the solid isochore "AB". However, when $E$ is high enough to exceed the maximum superheating, the system will transform into a state on the liquid isochore "CD". When this happens, some kinetic energy will convert into the potential energy of the liquid, leading to a lower temperature at point "C" (the lowest point on the liquid isochore) than at point "B" (the highest point on the solid isochore), thus, suppressing partially the superheating effects. Point "C" is taken as the melting point from the Z method. In this example, $T_m$ is given as $3720 \pm 20$ K, ~200 K (~6%) higher than that predicted by the more accurate, but also much more expensive coexistence method.

Accurate melting point simulation using the coexistence method begins from a half-solid and half-liquid initial configuration. We prepared this configuration following three steps: targeting at a temperature ($T_t$) ~10% lower than $T_m$ estimated by the Z method. First, a large supercell containing more than 600 atoms with a perfect crystalline structure was created, which is 4-5 times as long in one lattice direction (denoted by $z$ in the rest of this paper) as in the others. Such an elongated supercell helps to maintain solid-liquid coexistence during the simulation. Second, the temperature reduced gradually from a very high value (e.g., twice $T_t$) to $T_t$, to melt half of the system (at half of the long edge of the supercell), while the other half stayed frozen. Finally, the temperature rose gradually from a relatively low value (e.g., half $T_t$) to $T_t$, to heat the previously frozen half of the system to a high-temperature solid state, with the other half frozen instead. Starting from the prepared initial configuration, AIMD simulations were performed under the NVE ensemble, with different initial temperatures $T_0$ chosen to be within ~$\pm 20\%$ of $T_t$, corresponding to different internal energies $E$ of the system. At each volume, there is a range of $E$ at which solid-liquid coexistence can be maintained for a period of 10 ps or longer, and the melting point can be determined accurately as the pressure-temperature ($P$-$T$) condition of the equilibrium coexisting state. This is quite different from melting simulations performed under the NPT ensemble (constant number of atoms, constant pressure, and constant temperature), where it is difficult to maintain stable solid-liquid coexistence. In the NPT ensemble, since the considered $P$-$T$ condition generally does not lie on the melting curve, starting from an initial solid-liquid coexistence configuration, the system would always melt or solidify completely after some simulation time. As a result, the melting point cannot be determined directly under the NPT ensemble, instead a series of simulations have to be performed to locate the upper and lower bounds of the melting temperature.[47] For an efficient application in practice, we first performed an AIMD simulation with $T_0 = T_t$, then conducted another one only if the coexistence state was not maintained in the previous one. If the system completely solidifies/melts during the previous simulation, then $T_0$ will be raised/lowered for the next simulation.

The "block average" method is commonly employed to compute the standard error for a set of correlated data.[48] However, the direct application of this method requires that the dynamic quantity is measured for the same equilibrium phase during the simulation. In the context of solid-liquid coexistence simulations, there is in general a dynamic change in the proportions of solid and liquid phases in the system. As a result, the pressure and temperature measured are effectively the averages of a collection of different coexisting states. In this case, we find that using the standard "block average", the errors in pressure and temperature can be seriously underestimated. We therefore adopted a different method for error estimation. In analogous to the standard "block average", we first divide the data into $N$ blocks, with $M$ data in each block, and then calculate the average value for each block. $M$ is chosen to be large enough, so that the resulting $N$ block averages become uncorrelated to each other. In the end, the error is obtained as the variance of these block averages. The error determined as such is $\sqrt{N-1}$ times of that given by the standard "block average". In the present study, we find it appropriate to set $M = 200$ for all the coexistence simulations performed.

The volumes of solid and liquid phases are generally different, which can lead to nonhydrostatic stresses in coexistence simulations and influence the accuracy of the calculated melting points. The volume of the liquid is usually larger than that of the solid, except in the special case when reentrant melting occurs at high pressures, as shall be discussed in Sec. III for Mg. To check the degree of nonhydrostaticity in our coexistence simulations, we calculated the deviation of each normal stress component $P_i$ ($i = x, y, z$) from their average $\overline{P}$ as $\eta_i = (P_i - \overline{P})/\overline{P}$. Due to symmetry reasons, $\eta_x$ and $\eta_y$ are nearly the same and both about half of $-\eta_z$. In all our simulations where a stable solid-liquid coexistence state was maintained so that a melting point was obtained, $|\eta_i|$'s are all less than 1%, except at the lowest considered pressure (~7 GPa) where $\eta_z$ is -2.8% (due to the small value of $\overline{P}$ which makes $\eta_i$ sensitive). As a result, we believe that nonhydrostatic effects on the melting of Mg should not be important in the present work.

In MD simulations, in order to identify the instantaneous phase state of the system, the most straightforward way is the direct observation of atomic configuration, where an example is shown in Fig. 2. The ordered (colored in blue) and disordered (colored in red) parts of the configuration evidently correspond to solid and liquid structures, respectively. The order parameter (OP)[41, 42, 49] and the number density[41, 50, 51] provide alternative ways for quantitative identification of solid and liquid structures. Figure 3 shows the local OP and the number density calculated for Mg, both corresponding to the atomic configuration displayed in Fig. 2. The local OP[41, 49] of an atom is defined as

$$\psi = \left| \frac{1}{N_q} \frac{1}{Z} \sum_{\vec{r}} \sum_{\vec{q}} \exp(i\vec{q} \cdot \vec{r}) \right|^2, \tag{1}$$

where $\vec{r}$ represents a vector connecting this atom to one of its Z nearest neighbors found within a distance $r_c$, set according to the symmetry of the considered crystal structure,[41, 49] and each wave vector $\vec{q}$ was chosen to satisfy the relation $\exp(i\vec{q} \cdot \vec{r}) = 1$ for each $\vec{r}$ in the perfect crystal lattice. To reduce the large fluctuations in the local OP, we followed Morris and Song[49] to first average instantaneous atomic positions over short periods of time before calculating the local OP, then averaging the local OP with values of neighboring atoms. As can be seen by comparing Figs. 2 and 3(a), the resulting OP is very close to 0 for the liquid, which can be attributed to the disorder in the liquid structure, and is somewhere between 0 and 1 for the crystal. In fact, OP takes the value of 1 for a perfect crystal without the thermal vibrations of atoms. The number density is calculated by cutting the system into equally spaced slices perpendicular to the long edge of the supercell (direction z) and counting the number of atoms in each slice. As can be seen by comparing Figs. 2 and 3(b), the number density exhibits strong periodic oscillations in the solid part of the system, while oscillates only moderately in the liquid part.

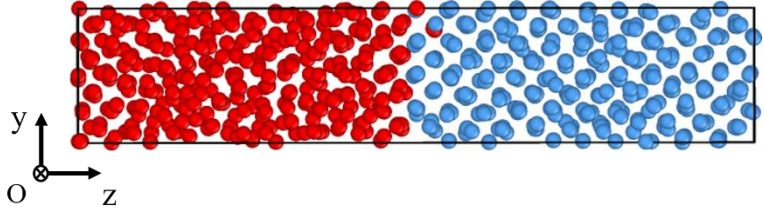

FIG. 2. Snapshot of an atomic configuration, taken in a solid-liquid coexistence simulation for the melting of the bcc phase of Mg at $V = 8.2$ Å$^3$/atom. The red and blue colors mark the liquid and solid phases, which correspond to more disordered and more ordered parts of the configuration, respectively.

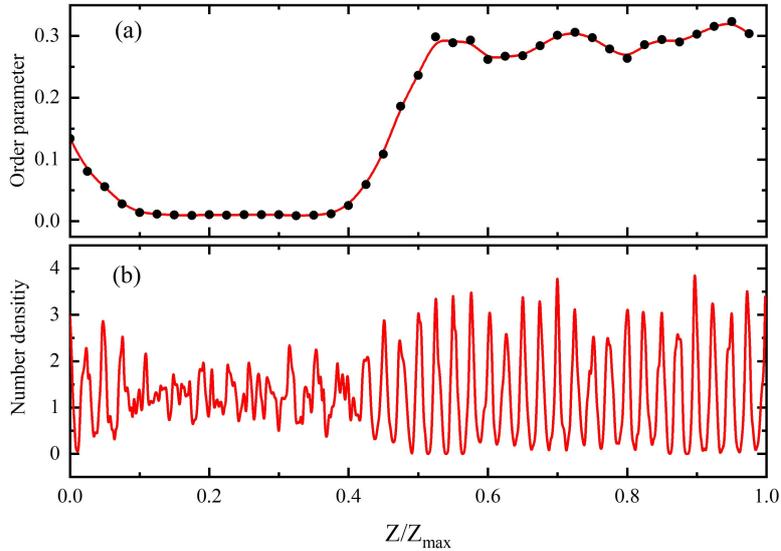

FIG. 3. Order parameter (a) and number density (b) profiles, corresponding to the atomic configuration displayed in Fig. 2.

## III. Results and Discussion

### A. Distinct characteristics in coexistence simulations of Mg

We find that when the solid-liquid coexistence method is applied under the NVE ensemble, at high pressures of a few hundred GPa, simulations of Mg show some characteristics distinctly different from other metal systems.[28, 29, 45] To reveal these characteristics, in the following we make detailed comparisons between examples for Mg and Al. In particular, we consider the example for the bcc phase of Mg at $V = 8.2$ Å$^3$/atom, in which case these characteristics are most pronounced, and the example for the fcc phase of Al at $V = 10.0$ Å$^3$/atom (in a $3 \times 3 \times 15$ supercell containing 540 atoms), as a typical representation of other metals. Four and five trial coexistence simulations were performed for Mg and Al, respectively, with the same initial solid-liquid coexistence configuration for the same metal, and the initial temperature $T_0$ in different simulations varying from 4000 to 5500 K for Mg, and from 5000 to 7000 K for Al, at an increment of 500 K. The pressure and temperature evolutions in these simulations (except for Al with $T_0 = 6000$ K) are reported in Figs. 4 and 5 for Mg and Al, respectively.

According to Sec. II, the phase state of the system during the simulation can be identified either by directly observing the atomic configuration, or from calculated values of the number density or the order parameter (OP). We find that in the example of Mg/Al, the system completely solidified after ~6 ps/~8 ps when $T_0$ was 4000 K/5000 K, completely melted after ~4 ps/~4 ps when $T_0$ was 5500 K/7000 K, and remained in a solid-liquid coexistence state after 10 ps when $T_0$ was {4500 K or 5000 K}/{5500 K, 6000 K or 6500 K}. From these simulations in which a coexistence state was maintained, two and three melting points were obtained for Mg and Al, respectively. They are [$282.5 \pm 0.2$ GPa, $4263 \pm 52$ K] and [$282.9 \pm 0.2$ GPa, $4310 \pm 60$ K] for Mg with $T_0 = 4500$ K and $T_0 = 5000$ K, respectively, and [$145.9 \pm 0.2$ GPa, $5151 \pm 65$ K] and [$149.6 \pm 0.1$ GPa, $5269 \pm 65$ K] for Al with $T_0 = 5500$ K and $T_0 = 6500$ K, respectively. Another melting point of Al, given by the simulation with $T_0 = 6000$ K, lies approximately in the middle of the other two melting points obtained for Al.

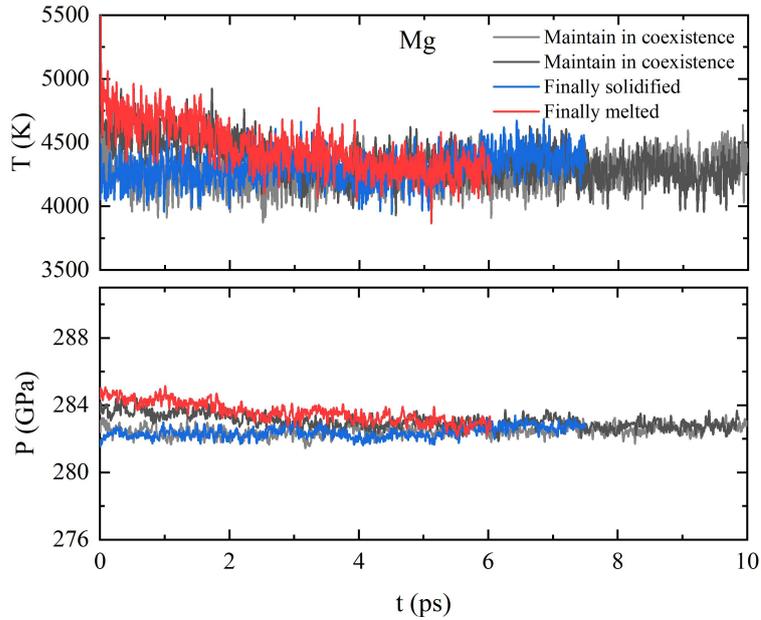

FIG. 4. Temperature (upper panel) and pressure (lower panel) evolution in solid-liquid coexistence simulations, for the bcc phase of Mg at $V = 8.2$ Å$^3$/atom. The blue, grey, black and red colors correspond to simulations with initial temperatures of 4000 K, 4500 K, 5000 K and 5500 K, respectively. The same initial atomic configuration was used in these simulations.

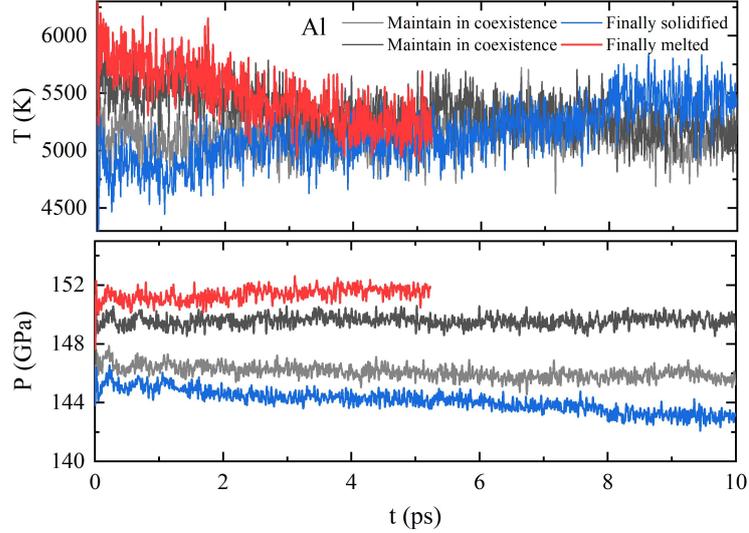

FIG. 5. Temperature (upper panel) and pressure (lower panel) evolution in solid-liquid coexistence simulations, for the fcc phase of Al at $V = 10.0$ Å$^3$/atom. The blue, grey, black and red colors correspond to simulations with initial temperatures of 5000 K, 5500 K, 6500 K and 7000 K, respectively. The same initial atomic configuration was used in these simulations.

As can be seen, the largest differences between the different melting points obtained for Mg/Al are ~0.4 GPa/~3.7 GPa in pressure and ~47 K/~118 K in temperature. These differences correspond to ~0.1%/~2.5% of the pressure and ~1.1%/~2.2% of the temperature at the calculated melting points of Mg/Al. Apparently, in these simulations, the pressure and temperature range where a stable coexistence state was found for Mg are much smaller than those for Al. The small pressure and temperature range for maintaining a solid-liquid coexistence state at a given volume is the first characteristic we find in coexistence simulations of Mg at high pressures.

Also note that the melting points obtained for Mg and Al are associated with a $T_0$ range of 500 and 1000 K, respectively. For the same metal, the difference in $T_0$ directly reflects the difference in internal energy $E$ (kinetic energy $K$ + potential energy $U$). This is because $T_0$ determines the initial $K$ ($K \equiv 3/2(N-1)k_BT$, where $N$ represents the number of atoms and $k_B$ denotes the Boltzmann constant), and the initial $U$ is the same in different simulations. Therefore, in these simulations, the internal energy range where a stable coexistence state was found for Mg is only half of that for Al. The relatively small internal energy range for maintaining a solid-liquid coexistence state at a given volume is a second characteristic we find in coexistence simulations of Mg at high pressures.

The simulation characteristics we find for Mg are closely related to the small pressure and internal energy differences between solid and liquid phases (at the same $V$-$T$ condition). To elaborate this point, in Fig. 6, we report for Mg and Al the calculated pressures and internal energies of both phases as a function of temperature, determined at the same volumes ($V = 8.2$ Å$^3$/atom for Mg, and $V = 10.0$ Å$^3$/atom for Al) as in the above coexistence simulations. These results were obtained from AIMD simulations performed in small supercells with around 100 atoms. The isochoric $P$-$T$ relations for Mg and Al, along with the corresponding melting curves, are presented in Figs. 6(a) and 6(b), respectively, and the isochoric $E$-$T$ relations for Mg and Al are presented in Figs. 6(c) and 6(d), respectively. The melting curve of Mg is from the present work (see Sec. III B), and that of Al is from a previous AIMD study.[52] In Fig. 6, we use "A"/"B" to represent the thermodynamic condition at the intersection of the melting curve with the solid/liquid $P$-$T$ isochore.

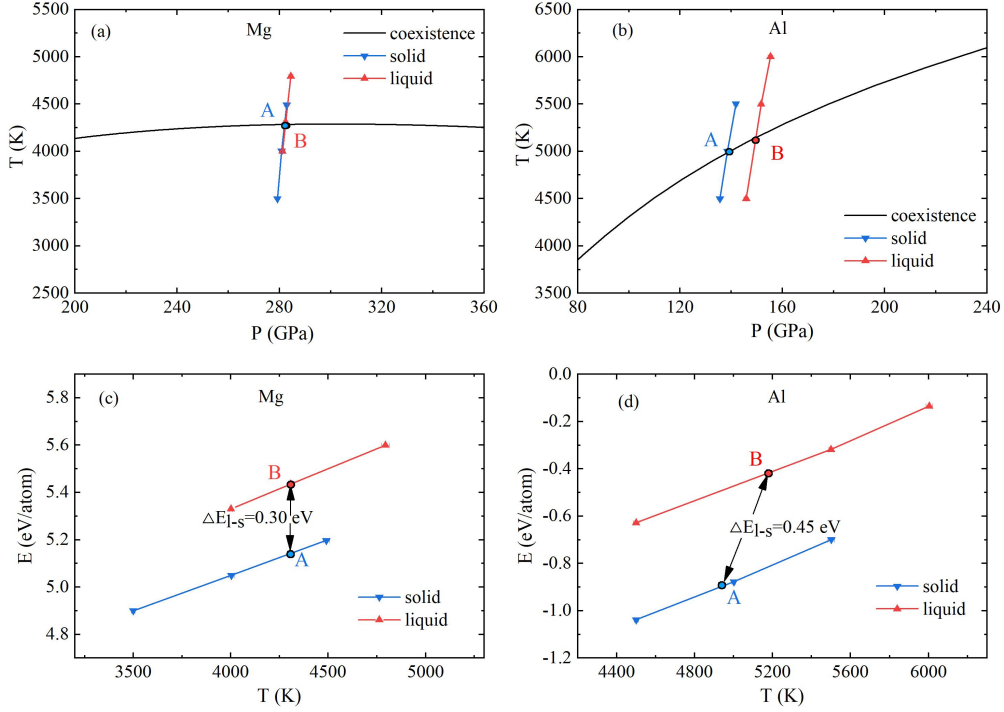

FIG. 6. Pressure (a) and internal energy (c) as a function of temperature on the bcc and liquid isochores of Mg at $V = 8.2$ Å$^3$/atom, and pressure (b) and internal energy (d) as a function of temperature on the fcc and liquid isochores of Al at $V = 10.0$ Å$^3$/atom. The melting curve of Mg in (a) is from the present work, and that of Al in (b) is from a previous AIMD study.[52] In these figures, "A" and "B" represent the thermodynamic conditions at the intersections of the melting curve with the solid and liquid $P$-$T$ isochores, respectively.

In a solid-liquid coexistence simulation under the NVE ensemble, the simulated system may end in three different states. In ideal situations without overheating and undercooling, when $E$ is lower/higher than the solid/liquid internal energy at condition "A"/"B" ($E_s^A$/$E_l^B$), the system would end in a state below/above the point "A"/"B" on the solid/liquid $P$-$T$ isochore [see Figs. 6(a) and 6(b)]. And, when $E$ is between $E_s^A$ and $E_l^B$, the system would end in a solid-liquid coexistence state between points "A" and "B" on the melting curve, with more solid/liquid phase if the internal energy is closer to $E_s^A$ / $E_l^B$ . Therefore, ideally, the pressure, temperature and internal energy range where solid-liquid coexistence can be maintained correspond to the pressure, temperature, and internal energy differences between the solid state at "A" and the liquid state at "B", respectively. As a result, these ranges are determined by the positions of solid and liquid $P$-$T$ isochores relative to the melting curve, and the differences in $P$-$T$ and $E$-$T$ relations between solid and liquid isochores. In practice, these ranges are affected by thermodynamic fluctuations and related to the duration of the simulation as well. Stronger thermodynamic fluctuations due to smaller system sizes and longer simulation time both tend to break the solid-liquid coexistence state.

As shown in Figs. 6(a) and 6(b), quite different from Al, in the example of Mg, the isochoric *P*-*T* relations are very close between solid and liquid phases, leading to a very small pressure and temperature range for maintaining solid-liquid coexistence. The internal energy range for maintaining solid-liquid coexistence in the example of Mg is also notably smaller, for two reasons evident from Figs. 6(c) and 6(d). First, under the same *V*-*T* condition, the internal energy difference between solid and liquid phases of Mg (~0.30 eV) is smaller than that of Al (~0.39 eV). Second, unlike Mg, in the example of Al, the temperature at condition "B" is noticeably higher than that at condition "A". This temperature difference (~160 K) further enlarges the internal energy difference between the solid state at condition "A" and the liquid state at condition "B", leading to a total internal energy difference of ~0.45 eV for Al, in comparison with a value of only ~0.30 eV in the example of Mg.

Through coexistence simulations at atomic volumes other than considered above, we confirm that the simulation characteristics revealed are common for Mg at high pressures of a few hundred GPa. This can be understood by comparing the solid-liquid pressure and internal energy differences between Mg and Al. These differences are reported as a function of atomic volume $V$ in Figs. 7(a) (pressure) and 7(c) (internal energy), and as a function of the corresponding liquid pressure $P_l(V)$ in Figs. 7(b) (pressure) and 7(d) (internal energy). In Fig. 7, red circles represent the results for Mg from our AIMD simulations, obtained in small supercells with around 100 atoms at temperatures around melting, and blue open circles and diamonds represent our static results for Mg and Al, respectively. These static results were obtained using a simple scaling method following two steps, as quick approximations to AIMD results.[14] First, a random snapshot is taken from an AIMD simulation, to represent the appropriate phase state, either solid or liquid. Second, static calculations are performed at a series of atomic volumes, with atomic configurations obtained by uniformly scaling the chosen snapshot in all three lattice directions. As can be seen in Fig. 7(a), the solid-liquid pressure difference of Mg from our static calculations is in good agreement with those calculated by Hong and van de Walle[14] using the same method (grey solid line). Moreover, as shown in Figs. 7(a)-(d), for Mg, the variation tendencies of the calculated solid-liquid pressure and internal energy differences at different compressions are quite similar between static and AIMD results, indicating that the simple scaling method we have employed works effectively at conditions around melting. As shown by the static results in Figs. 7(b) and 7(d), at high compressions, the solid-liquid pressure and internal energy differences of Mg are both notably smaller than those of Al. In particular, as $P_l(V)$ increases from ~100 to 500 GPa, the pressure difference $P_l - P_s$ of Mg decreases from ~3 to -5 GPa, while that of Al varies between ~9 and 12 GPa. And, as $P_l(V)$ increases from ~150 to 500 GPa, the internal energy difference $E_l - E_s$ of Mg remains almost constant at ~0.3 eV, while that of Al increases from ~0.4 to 0.6 eV. According to previous analyses, the small pressure and internal energy differences of Mg would lead directly to the two simulation characteristics discussed above, i.e., the small pressure and temperature range, as well as the small internal energy range, for maintaining solid-liquid coexistence.

These small differences between solid and liquid phases of Mg can be attributed to the "uneven" softening of interatomic interactions, which occurs faster in the liquid than in the solid as the structure is continuously compressed.[14,53] In Fig. 7(b), we can see that as pressure increases, faster softening in the liquid phase of Mg starts at ~100 GPa. The effects are so strong that $P_l - P_s$ of Mg decreases rapidly as pressure further increases, and becomes very small and even negative at high pressures. With the strong softening effects, the liquid is in a more relaxed state compared with the case without it. This effectively lowers the liquid internal energy, making $E_l - E_s$ of Mg small at high pressures, as shown in Fig. 7(d). In comparison, faster softening in the liquid phase of Al starts at a much higher pressure of ~300 GPa, and the effects are also much less pronounced. The results are that $P_l - P_s$ of Al decreases by less than 2 GPa as pressure increases from ~300 to 500 GPa, as shown in Fig. 7(b), and at high pressures $E_l - E_s$ of Al exhibits a much stronger pressure dependence than that of Mg, as shown in Fig. 7(d).

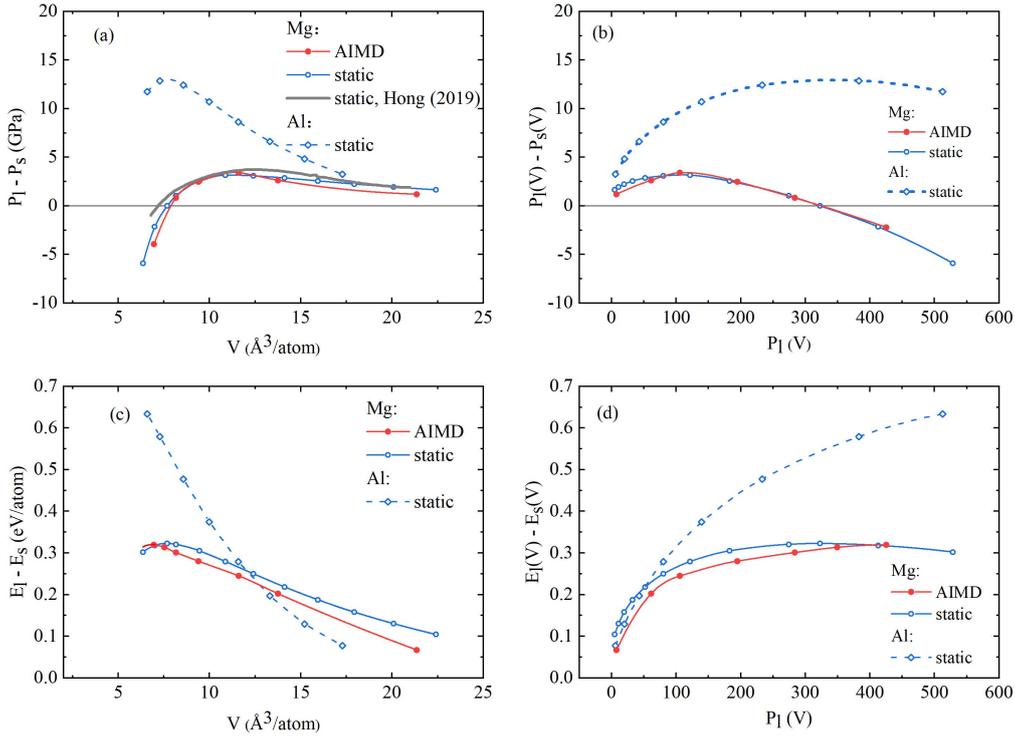

FIG. 7. Pressure difference between liquid and solid phases as a function of volume (a) and the corresponding liquid pressure (b) for Mg and Al, and internal energy difference between liquid and solid phases as a function of volume (c) and the corresponding liquid pressure (d) for Mg and Al. Red and blue colors represent our calculated results (see text), and grey color denotes the results from Hong and van de Walle.[14]

The stronger potential softening in the liquid phase also leads to the occurrence of reentrant melting for Mg at high pressures, an unusual behavior that the melting temperature declines with increasing pressure, i.e., the negative slope of the melting curve.[14] According to the Clausius-Clapeyron relation, the slope of the melting curve is determined by

$$\frac{dT_m}{dP} = \frac{\Delta V}{\Delta S}, \quad (2)$$

where $\Delta V = V_l - V_s$ and $\Delta S = S_l - S_s$ represent the volume and entropy differences between liquid and solid phases at $P$-$T$ conditions of melting, respectively. $\Delta S$ is always greater than zero, since $S_l$ is always larger than $S_s$ due to structural disorder in the liquid. As a result, the sign of the slope of the melting curve is determined by $\Delta V$. In normal cases, $V_l > V_s$ so that $\Delta V > 0$, leading to a positive slope of the melting curve. However, as can be seen from Figs. 7(a) and 7(b), due to the strong potential softening in the liquid phase, $P_l$ becomes smaller than $P_s$ at small volumes, corresponding to $P > \sim 320$ GPa. This equivalently means that at $P > \sim 320$ GPa $V_l < V_s$ so that $\Delta V$ becomes less than 0, leading to a negative slope of the melting curve, i.e., reentrant melting. The pressure for the onset of reentrant melting as inferred from Fig. 7(b) (~320 GPa) agrees with that obtained by Hong and van de Walle based on similar analyses (~300 GPa).[14] Using the simple scaling method discussed above, Hong and van de Walle[14] estimated $\Delta V$ as a function of pressure for some other metals as well, and speculated that reentrant melting is a universal feature of all metals. However, they predicted that for many metals the onset of this behavior is at a much higher pressure, e.g., ~3500 GPa for Al.[14] One exception is sodium (Na), for which experiments found reentrant melting at $P > \sim 30$ GPa.[54] Considering that reentrant melting is also caused by the strong potential softening in the liquid phase, we believe that the simulation characteristics we found in the present work for Mg at high pressures should also be present for other metals at pressures near the onset of reentrant melting.

We find that in coexistence simulations under the NVE ensemble, at high pressures, Mg shows another characteristic distinctively different from other metal systems, such as Al. For Al, when the system continues to melt or solidify, the variation tendencies of pressure and temperature are opposite, as can be seen from Fig. 5. This is the typical behavior exhibited by other metal systems.[28, 29, 45] However, Fig. 4 shows that when the same happens for Mg, pressure and temperature tend to vary in the same direction, i.e., both increasing or decreasing. This different behavior is also related to the small solid-liquid pressure difference of Mg. Consider when the system continues to melt. In this case, the potential energy of the system tends to increase, since the solid transforms into the liquid with higher potential energy. Because the total internal energy remains constant, the kinetic energy must decrease to compensate for the increase of potential energy, leading to a drop in temperature, no matter which metal is being considered. Although the temperature drop itself tends to decrease pressure, how the system pressure varies depends on the solid-liquid pressure difference $P_l - P_s$ as well. For Al, $P_l - P_s$ is relatively large [see Fig. 7(b)], and, therefore, more liquid phase due to melting becomes a strong driving force to increase the system pressure. However, for Mg, $P_l - P_s$ can be very small at pressures around the onset of reentrant melting or even negative at higher pressures. In this case, the net effects are that the system pressure tends to decrease so that pressure and temperature vary in the same direction. The solidification is the reverse process and can be analyzed similarly. One may suspect that the abnormal tendency for pressure and temperature to vary in the same direction is related to the energy drift in MD simulations. We have checked to make sure that this abnormal behavior is present also in simulations with slower energy drift, which is achieved by simultaneously reducing the time step and the total energy convergence threshold. The energy drift is of no significance here, since different from the cases when the system continues to melt or solidify, during a short simulation period the energy drift is not fast enough to cause noticeable pressure and temperature variations.

The last characteristic we would like to discuss is the relatively small changes in pressure and temperature levels (referring to the value around which the instantaneous values fluctuate) when phase transition occurs in coexistence simulations of Mg at high pressures. As shown in Fig. 4, as Mg evolves from the initial solid-liquid coexistence state to complete melting or solidification, the pressure level changes by ~0.5-1 GPa, and the temperature level changes by ~150-300 K. However, Fig. 5 shows that for Al the variations are noticeably larger, ~1-2 GPa and ~500-700 K. The smaller pressure variation for Mg is because the solid-liquid pressure difference of Mg is relatively small, and the phase transition does not affect the pressure as evidently as in the case of Al. The smaller temperature variation for Mg is because the solid-liquid internal energy difference of Mg is relatively small, and upon phase transition, less amount of energy needs to be converted between the kinetic energy and the potential energy.

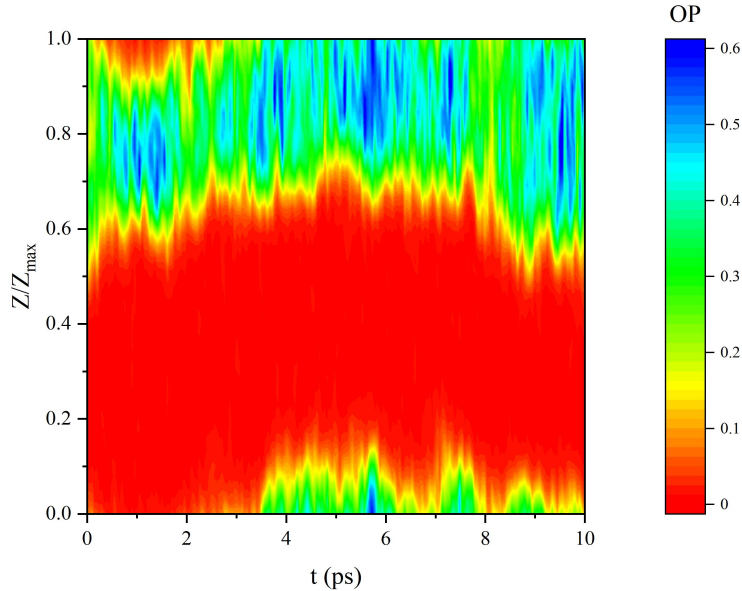

FIG. 8. Could diagram for the OP profile, from the solid-liquid coexistence simulation for the melting of the bcc phase of Mg at $V = 8.2$ Å$^3$/atom with the initial temperature of 4500 K.

Due to the relatively small variations in pressure and temperature levels during phase transition, whether the system is melting or solidifying in a coexistence simulation of Mg cannot be easily identified according to the pressure and temperature evolutions such as shown in Fig. 4. To monitor the phase state of the system more conveniently, a cloud diagram displaying the time evolution of the OP profile can be plotted, where an example is given in Fig. 8. In this figure, colors are used to represent the values of the OP. The region with OP's close to 0 (colored in red) indicates a disordered structure, i.e., the liquid, while the region with OP's obviously larger than 0 (colored in green and blue) indicates an ordered structure, i.e., the solid. As shown in Fig. 8, such a cloud diagram does not only clearly illustrate the time evolution of solid and liquid proportions in the system, but also reveals that even when the coexistence state is maintained at the end of the simulation, solid-liquid interfaces can move somehow (forward and backward) during the simulation.

## B. Melting curve of Mg

We conducted solid-liquid coexistence simulations at nine atomic volumes based on $4 \times 4 \times 20$ supercells (640 atoms), and consequently obtained nine melting points for the bcc phase of Mg, at pressures between ~7 and 460 GPa. Our results are compared with previous experimental measurements and first-principles calculations in Fig. 9, and data of the calculated melting points are presented in Table I. The hcp-bcc phase boundaries from previous first-principles calculations[6, 8, 10] and static measurements[13] are also shown in the figure, which serves as references for the stability domains of hcp and bcc phases. To check the convergence of our calculated melting points, we performed a series of test simulations at ~105 and ~460 GPa based on different supercell sizes, where each simulation lasted for 8-10 ps. In Fig. 9, the melting points from some of these simulations are compared with the results from $4 \times 4 \times 20$ supercells, based on which our melting curve is fitted. At ~460 GPa, the melting temperature from the $4 \times 4 \times 14$ supercell (448 atoms) is $4020 \pm 60$ K, which agrees excellently with the value of $4039 \pm 47$ K from the $4 \times 4 \times 20$ supercell. Using $5 \times 5 \times 14$ (700 atoms) supercells, which are longer in both the x and y directions, the melting temperatures at ~105 GPa and ~460 GPa are 95 K (2.7%) and 184 K (4.6%) lower than the values from $4 \times 4 \times 20$ supercells, respectively. This indicates that size effects are stronger at higher pressures, when Coulomb interactions become stronger due to smaller interatomic distances. Using an even larger $6 \times 6 \times 14$ (1008 atoms) supercell at ~460 GPa, the melting temperature is 133 K (3.3%) lower than the value from the $4 \times 4 \times 20$ supercell but agrees with that from the $5 \times 5 \times 14$ supercell within computational uncertainty. To check if the **k**-point sampling has a significant influence, we performed a test simulation of 5 ps at ~460 GPa using a $2 \times 2 \times 1$ **k**-point grid for the $4 \times 4 \times 20$ supercell, which is denser in the x and y directions. The resulting melting point of [463.5 ± 0.7 GPa, 4073 ± 67 K] agrees with that calculated using the $\Gamma$-only **k**-point grid within computational uncertainty, although the average temperature is slightly higher by 34 K. This seems to indicate that in the present study the calculated melting point does not depend sensitively on the **k**-point sampling. The above comparisons show that the convergence of our calculated melting temperature is probably better than 3% (at most ~100 K) below ~105 GPa, where experimental data are available, and better than 5% (at most ~200 K) at higher pressures up to ~460 GPa.

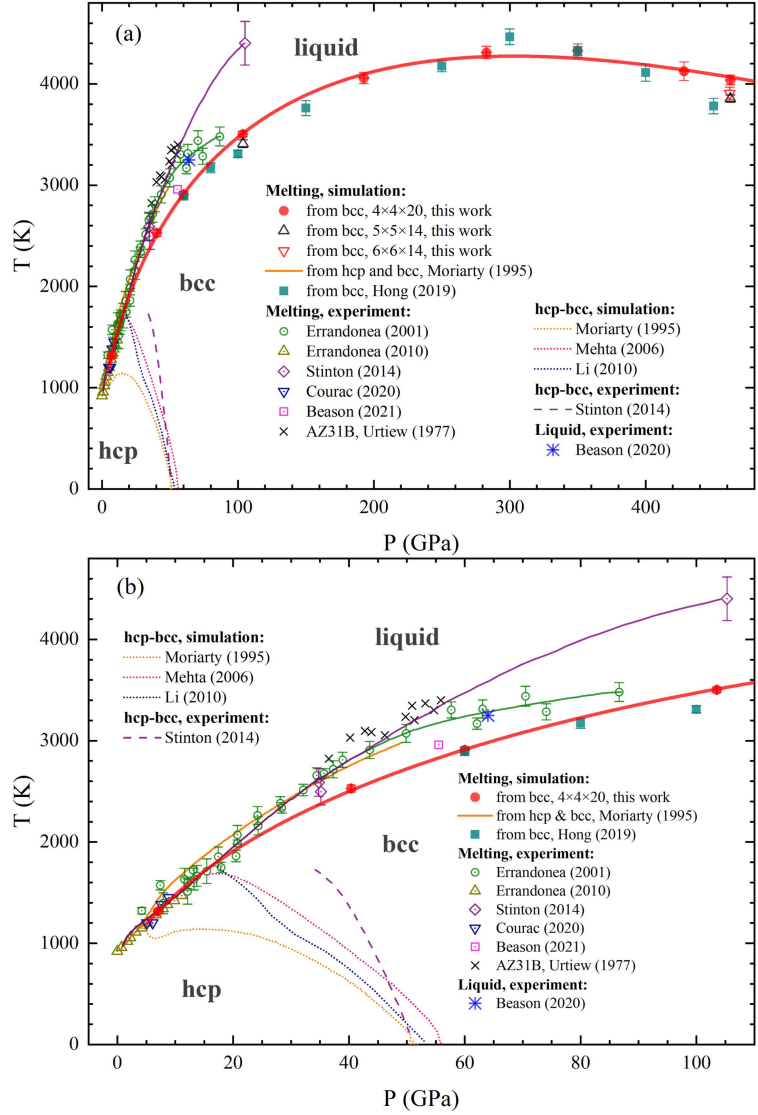

FIG. 9. Our results for the melting of Mg calculated using the solid-liquid coexistence method, and comparison with previous experimental measurements and first-principles calculations. Subfigures (a) and (b) correspond to different pressure ranges, up to 480 GPa and 110 GPa, respectively. Experimental and theoretical melting points are marked with open and solid symbols, respectively. Red solid circles represent our calculated melting points for the bcc phase of Mg, determined using $4 \times 4 \times 20$ supercells, and the red line is the corresponding fit to a low-order Kechin equation[30] (see text); up and down empty triangles denote our melting points calculated using $5 \times 5 \times 14$ and $6 \times 6 \times 14$ supercells, respectively. Green open circles and green solid line: static experiments by Errandonea et al.;[11] purple open diamonds and purple solid line: static experiments by Stinton et al.;[13] yellow open triangles: static experiments by Errandonea;[12] blue open triangles: static experiments by Courac et al.;[15] pink open squares: shock experiments by Beason et al.;[16] green solid squares: first-principles calculations by Hong and van de Walle[14] using the small-cell coexistence method; orange solid line: first-principles calculations by Moriarty and Althoff[8] using the free energy method. hcp-bcc phase boundaries from previous first-principles calculations (dotted lines)[6, 8, 10] and static measurements (dashed line)[13] are also shown, along with the melting points of AZ31B (crosses),[56] and a liquid state point from the shock experiments by Beason et al. (star, with temperature estimated by our AIMD simulations).[23]

TABLE I. Melting points for the bcc phase of Mg, calculated at different atomic volumes using the solid-liquid coexistence method, based on 4 × 4 × 20 supercells.

| $V$ (Å$^3$/atom) | $P$ (GPa) | $T$ (K) |
|---|---|---|
| 21.36 | 7.0 ± 0.1 | 1315 ± 20 |
| 15.33 | 40.4 ± 0.1 | 2528 ± 36 |
| 13.76 | 60.0 ± 0.1 | 2909 ± 32 |
| 11.61 | 103.5 ± 0.1 | 3504 ± 27 |
| 9.39 | 192.7 ± 0.1 | 4059 ± 54 |
| 8.17 | 282.9 ± 0.2 | 4310 ± 60 |
| 7.53 | 350.0 ± 0.3 | 4329 ± 67 |
| 6.96 | 428.3 ± 0.9 | 4126 ± 91 |
| 6.75 | 462.4 ± 0.6 | 4039 ± 47 |

Due to the unusual reentrant melting behavior of Mg at high pressures, our calculated melting points cannot be fitted using the commonly employed Simon-Glatzel equation[55] $T_m = T_0(1 + P/a)^b$, which can only describe melting curves with a positive slope. Kechin proposed a set of equations to describe different behaviors for the melting temperature dependence on pressure, rising, falling, flattening, as well as with a maximum.[30] We find that our calculated melting points can be excellently fitted to a low-order form of the Kechin equation[30] $T_m = T_0(1 + P/a)^b \exp(-cP)$, which is effectively a simple modification of the Simon-Glatzel equation. With the additional exponentially decaying term $\exp(-cP)$, this equation leads to a negative slope of the melting curve at high pressures, making it possible for us to obtain a melting curve of Mg at high pressures up to ~460 GPa. The fitted parameters are $T_0 = 782$ K, $a = 3.5796$ GPa, $b = 0.4905$, and $c = 0.0016$ GPa$^{-1}$. Since hcp is the more phase stable at relatively low pressures, it is reasonable that $T_0$ for the bcc phase is lower than the ambient-pressure experimental value of 921 K for the hcp phase. The pressure (~305 GPa) for the onset of reentrant melting given by the fitted equation, as the pressure at which $dT_m/dP$ is zero, is close to the pressure (~320 GPa) at which the atomic volumes of solid and liquid phases become equal, as determined in Sec. III A at conditions around melting.

As shown in Fig. 9, experiments on the melting of Mg have only been conducted up to 105 GPa. Between ~10 and 20 GPa, our calculated melting curve for the bcc phase is in reasonable agreement with static experiments,[11, 12, 15] though it is controversial whether Mg melts from the hcp or the bcc phase in this pressure regime.[6, 8, 10, 13] Between ~30 and 80 GPa, the melting temperatures from the static experiments by Errandonea *et al.*[11] are systematically higher than our results, by ~300 K in average. Stinton *et al.*[13] obtained a melting curve from static experiments that is even higher at pressures above ~50 GPa. At 105 GPa, the highest pressure reached in their experiments, the measured melting temperature (~4400 K) is ~800 K higher than the result (~3600 K) extrapolated from Errandonea *et al.*'s melting curve[11] and ~900 K higher than our result (~3500 K). The large discrepancies at high pressures between different static experiments may be related to the different melting criterions adopted.

In the study of Stinton et al.,[13] melting was identified according to the structure inside the sample, but the temperature was measured from the surface which was heated. The temperature may be higher on the surface than in the inner part, leading to some overestimation in the measured melting temperature, as pointed out by Hong and van de Walle.[14] In the study of Errandonea et al.,[11] melting was identified using the speckle method by observing movement at the sample surface. This observation may become difficult at high pressures when the viscosity of the liquid is high, resulting in some uncertainty in the measured melting temperature, as indicated by Stinton et al.[13]

In two recent studies based on shock experiments, Beason et al. first used dynamic x-ray diffraction (XRD) to locate a melting point at $P = 55 \pm 2$ GPa,[23] then performed sound speed measurements to find incipient melt at $P = 55.5 \pm 0.3$ GPa,[16] These two pressures for shock melting are in good consistency with each other. In the former study by Beason et al.,[23] the temperature at the melting point was estimated to be between 2550 and 3120 K based on two EOS models, SESAME 2860 and Mie-Grüneisen. In their latter study,[16] various experimental data were used to calibrate a multiphase EOS, which is expected to provide better constraints on the temperature. From this EOS, the temperature at incipient melt was estimated as 2960 K, which lies within the temperature range given by their former study. It is worth noting that, although at pressures above ~30 GPa our melting curve is noticeably lower than those from the static experiments by Errandonea et al.[11] and Stinton et al.,[13] at $P = 55.5$ GPa, our melting temperature of ~2850 K is in good agreement with the value of 2960 K from the latter shock study by Beason et al.[16]

There are also some other experimental data related to the melting of Mg. In the shock experiments by Beason et al., a pure liquid state for Mg was detected at $P = 63$ GPa.[23] Based on the measured pressure and volume, we performed AIMD simulations to estimate the corresponding temperature, which supposedly provides an upper bound for the melting temperature. The estimated value of ~3250 K is higher than our melting temperature of ~3000 K at the same pressure, as can be seen in Fig. 9. However, it is comparable to the melting temperatures at similar pressures from the static experiments by Errandonea et al.,[11] and is even lower by ~200 K than the melting curve from the static experiments by Stinton et al.,[13] thus in stark contradiction to the results of Stinton et al. Due to lack of experimental data for Mg, the shock melting points measured by Urtiew and Grover for AZ31B,[56] an alloy in which Mg accounts for ~96% of the total weight, were often compared in the literature with the melting of Mg. As shown in Fig. 9, at pressures between ~30 and 60 GPa, the melting points of AZ31B are slightly higher than the melting curves of Mg from Errandonea et al.[11] and Stinton et al.,[13] by only around 100 K, but are much higher than our melting curve of Mg, by ~500 K in average. These comparisons between AZ31B and Mg show that, to clarify how compositions influence the melting points of Mg alloys, accurate knowledge of the melting of Mg is necessary.

Only two first-principles studies have investigated the melting of Mg before our present work. In an early study by Moriarty and Althoff,[8] the phase diagram of Mg was determined using the free energy method, where the melting curve was calculated up to 50 GPa. For the solid hcp and bcc phases, the free energy was determined using the quasi-harmonic approximation (QHA), which neglects the effects of lattice anharmonicity that might be important at high temperatures.[18, 19] And for the liquid phase, the free energy was determined using variational perturbation theory, which unlike the formally exact thermodynamic integration (TI) method[26, 27, 57] often employed nowadays, yields some approximate values for the liquid free energy. As shown in Fig. 9, their melting curve is overall consistent with static experiments,[11, 12, 15] although slightly higher than the static results from Errandonea et al.[11] at pressures between ~10 and 20 GPa. As pressure rises, their melting curve becomes increasingly higher than ours, with a maximum difference of ~200 K at $P = 50$ GPa.

Very recently, Hong and van de Walle[14] calculated the melting points for the bcc phase of Mg from 60 to 450 GPa using a statistical method they developed, the small-cell coexistence method, but did not provide the corresponding melting curve. In this method, MD simulations are performed in small supercells with ~100-200 atoms starting from an initial solid-liquid coexistence configuration, and the melting temperature is extracted by analyzing the probability distribution for the final state (solid or liquid) of the system. As shown in Fig. 9, up to ~400 GPa, the results of Hong and van de Walle are generally consistent with ours, with only one melting point slightly higher than our melting curve by ~160 K, and the others lower by no more than 120 K. There is a relatively large discrepancy at ~450 GPa, where their melting point is lower by ~230 K. As we have tested, using a larger $6 \times 6 \times 14$ supercell (1008 atoms) will lower our melting temperature by ~130 K, thus bringing our result closer to the results of Hong and van de Walle. According to previous investigations on other systems, simulations performed in small supercells tend to overestimate the melting temperature.[57, 58] For example, Sun et al.[57] reported that the high-pressure melting temperature of the iron can be overestimated by as much as ~10% using a small supercell of ~100 atoms. Despite the small supercells employed by Hong and van de Walle, most of their melting points show good consistency with our melting curve. This may be attributed to the optimal design of supercell shape in the small-cell coexistence method, which minimizes the interaction among periodic images, and hence the finite-size effects for the chosen supercell size.[59, 60] Note also that our calculated melting temperature varies smoothly with pressure, while the melting points from Hong and van de Walle[14] seem to be more scattered, and as a result difficult to be well represented by a smooth function such as a low-order Kechin equation.[30] In comparison with experiments, like our calculations, below ~100 GPa, the results of Hong and van de Walle[14] are also noticeably lower than the melting curves from the static experiments by Errandonea et al.[11] and Stinton et al.,[13] but are also in good agreement with the shock experiments by Beason et al.[16] Moreover, Hong and van de Walle[14] also predicted that reentrant melting occurs for Mg at very high pressures, and the onset of reentrant melting (~300 GPa) agrees with that (~305 GPa) from our calculations.

## IV. CONCLUSIONS

In this work, we performed AIMD simulations to determine the melting curve of the bcc phase of Mg up to ~460 GPa using the solid-liquid coexistence method. Between ~30 and 100 GPa, our melting curve is noticeably lower than those from the static experiments by Errandonea *et al.*[11] and Stinton *et al.*,[13] and that calculated using the free energy method by Moriarty and Althoff,[8] but is in good agreement with the results from recent shock experiments by Beason *et al.*[16, 23] Between ~60 and 450 GPa, our melting curve is overall consistent with the melting points calculated using the small-cell coexistence method by Hong and van de Walle.[14] To fully resolve the controversies in the high-pressure melting of Mg, further experimental investigations are necessary.

Our calculations show that, at very high pressures, Mg exhibits the unusual behavior of reentrant melting, that is, the melting temperature decreases as pressure further rises. The predicted onset of reentrant melting (~305 GPa) is in good agreement with that calculated by Hong and van de Walle (~300 GPa).[14] Due to this unusual behavior, our calculated melting points cannot be fitted using the commonly employed Simon-Glatzel equation,[55] which can only describe melting curves with a positive slope. However, we find that they can be excellently fitted to a low-order form of the Kechin equation,[30] making it possible for us to obtain a first-principles melting curve of Mg at high pressures above 50 GPa for the first time.

We find that in solid-liquid coexistence simulations under the NVE ensemble, at high pressures of a few hundred GPa, Mg shows some characteristics distinctively different from other metal systems, such as Al. First, the pressure and temperature range for maintaining solid-liquid coexistence can be very small, and the corresponding internal energy range is also relatively small. Second, when phase transition occurs, pressure and temperature tend to vary in the same direction, unlike other metals where the variations are in opposite directions. Third, the changes in pressure and temperature levels due to phase transition are relatively small. These characteristics can all be attributed to the stronger softening of interatomic interactions in the liquid than in the solid at high pressures, which makes the pressure and internal energy differences between solid and liquid phases of Mg relatively small. The strong softening in the liquid phase is also the origin of the reentrant melting behavior of Mg at high pressures.

This work provides useful references for resolving the controversies in the high-pressure melting of Mg. In addition, the analyses for the simulation characteristics exhibited by Mg facilitate a better understanding of melting calculations at high pressures. Similar simulation characteristics, as well as reentrant melting, are also expected for other systems with strong softening in the liquid phase at high pressures.

## ACKNOWLEDGEMENTS

This work was supported by the Foundation of LCP (Grant No. 6142A05210404) and the National Natural Science Foundation of China (NNSFC) (Grant Nos. 91730302, 11501039, and U1804123). Part of the simulations were performed on the

## AUTHOR DECLARATIONS

### Conflict of Interest

The authors have no conflicts to disclose.

## DATA AVAILABILIT

The data that support the findings of this study are available from the corresponding authors upon reasonable request.

## REFERENCES

[1]Z. Jin, P. Gumbsch, K. Lu, and E. Ma, Phys. Rev. Lett. **87**, 055703 (2001).
[2]D. Alfè, C. Cazorla, and M. J. Gillan, J. Chem. Phys. **135**, 024102 (2011).
[3]A. Samanta, M. E. Tuckerman, T. Q. Yu, and W. E, Science **346**, 729 (2014).
[4]H. F. Liu, H. F. Song, Q. L. Zhang, G. M. Zhang, and Y. H. Zhao, Matter Radiat. Extremes **1**, 123 (2016).
[5]H. F. Song and H. F. Liu, Phys. Rev. B **75**, 245126 (2007).
[6]S. Mehta, G. D. Price, and D. Alfè, J. Chem. Phys. **125**, 194507 (2006).
[7]G. Robert, P. Legrand, and S. Bernard, Phys. Rev. B **82**, 104118 (2010).
[8]J. A. Moriarty and J. D. Althoff, Phys. Rev. B **51**, 5609 (1995).
[9]D. Errandonea, Y. Meng, D. Häusermann, and T. Uchida, J. Phys. Condens. Matter **15**, 1277 (2003).
[10]P. F. Li, G. Y. Gao, Y. C. Wang, and Y. M. Ma, J. Phys. Chem. C **114**, 21745 (2010).
[11]D. Errandonea, R. Boehler, and M. Ross, Phys. Rev. B **65**, 012108 (2001).
[12]D. Errandonea, J. Appl. Phys. **108**, 033517 (2010).
[13]G. W. Stinton, S. G. MacLeod, H. Cynn, D. Errandonea, W. J. Evans, J. E. Proctor, Y. Meng, and M. I. McMahon, Phys. Rev. B **90**, 134105 (2014).
[14]Q. J. Hong and A. van de Walle, Phys. Rev. B **100**, 140102 (2019).
[15]A. Courac, Y. Le Godec, V. L. Solozhenko, N. Guignot, and W. A. Crichton, J. Appl. Phys. **127**, 055903 (2020).
[16]M. T. Beason, B. J. Jensen, and S. D. Crockett, Phys. Rev. B **104**, 144106 (2021).
[17]H. Olijnyk and W. B. Holzapfel, Phys. Rev. B **31**, 4682 (1985).
[18]J. W. Xian, J. Yan, H. F. Liu, T. Sun, G. M. Zhang, X. Y. Gao, and H. F. Song, Phys. Rev. B **99**, 064102 (2019).
[19]F. Soubiran and B. Militzer, Phys. Rev. Lett. **125**, 175701 (2020).
[20]A. K. McMahan and J. A. Moriarty, Phys. Rev. B **27**, 3235 (1983).
[21]G. R. Chavarría, Phys. Lett. A **336**, 210 (2005).
[22]C. J. Pickard and R. J. Needs, Nat. Mater. **9**, 624 (2010).
[23]M. T. Beason, A. Mandal, and B. J. Jensen, Phys. Rev. B **101**, 024110 (2020).
[24]J. R. Morris, C. Z. Wang, K. M. Ho, and C. T. Chan, Phys. Rev. B **49**, 3109 (1994).

[25]A. V. Karavaev, V. V. Dremov, and T. A. Pravishkina, Comput. Mater. Sci. **124**, 335 (2016).

[26]D. Alfe, G. D. Price, and M. J. Gillan, Phys. Rev. B **65**, 165118 (2002).

[27]L. Vočadlo and D. Alfè, Phys. Rev. B **65**, 214105 (2002).

[28]D. Alfè, Phys. Rev. B **68**, 064423 (2003).

[29]D. Alfè, Phys. Rev. B **79**, 060101 (2009).

[30]V. V. Kechin, Phys. Rev. B **65**, 052102 (2001).

[31]G. Kresse and J. Hafner, Phys. Rev. B **47**, 558 (1993).

[32]G. Kresse and J. Furthmüller, Comput. Mater. Sci. **6**, 15 (1996).

[33]G. Kresse and J. Furthmüller, Phys. Rev. B **54**, 11169 (1996).

[34]W. Kohn and L. J. Sham, Phys. Rev. **140**, A1133 (1965).

[35]P. E. Blöchl, Phys. Rev. B **50**, 17953 (1994).

[36]G. Kresse and D. Joubert, Phys. Rev. B **59**, 1758 (1999).

[37]J. P. Perdew, K. Burke, and M. Ernzerhof, Phys. Rev. Lett. **77**, 3865 (1996).

[38]K. Schwarz, J. Solid State Chem. **176**, 319 (2003).

[39]P. Blaha, K. Schwarz, G. K. Madsen, D. Kvasnicka, and J. Luitz, WIEN2k: An augmented plane wave plus local orbitals program for calculating crystal properties, Institute of Physical and Theoretical Chemistry, TU Vienna **60**, (2001).

[40]N. D. Mermin, Phys. Rev. **137**, A1441 (1965).

[41]S. C. Wang, H. F. Liu, G. M. Zhang, and H. F. Song, J. Appl. Phys. **114**, 163514 (2013).

[42]S. C. Wang, G. M. Zhang, H. F. Liu, and H. F. Song, J. Chem. Phys. **138**, 134101 (2013).

[43]A. B. Belonoshko, N. V. Skorodumova, A. Rosengren, and B. Johansson, Phys. Rev. B **73**, 012201 (2006).

[44]A. B. Belonoshko, L. Burakovsky, S. P. Chen, B. Johansson, A. S. Mikhaylushkin, D. L. Preston, S. I. Simak, and D. C. Swift, Phys. Rev. Lett. **100**, 135701 (2008).

[45]T. T. Zhang, S. C. Wang, H. F. Song, S. Q. Duan, and H. F. Liu, J. Appl. Phys. **126**, 205901 (2019).

[46]J. Bouchet, F. Bottin, G. Jomard, and G. Zérah, Phys. Rev. B **80**, 094102 (2009).

[47]D. Alfè, M. Gillan, and G. Price, J. Chem. Phys. **116**, 6170 (2002).

[48]D. Frenkel and B. Smit, *Understanding molecular simulation: from algorithms to applications*. Vol. 1. 2001: Elsevier.

[49]J. R. Morris and X. Song, J. Chem. Phys. **116**, 9352 (2002).

[50]Z. L. Liu, L. C. Cai, X. R. Chen, and F. Q. Jing, Phys. Rev. B **77**, 024103 (2008).

[51]Z. Y. Zeng, C. E. Hu, L. C. Cai, X. R. Chen, and F. Q. Jing, J. Appl. Phys. **109**, 043503 (2011).

[52]Z. L. Liu, X. L. Zhang, and L. C. Cai, J. Chem. Phys. **143**, 114101 (2015).

[53]J. Y. Raty, E. Schwegler, and S. A. Bonev, Nature **449**, 448 (2007).

[54]E. Gregoryanz, O. Degtyareva, M. Somayazulu, R. J. Hemley, and H. K. Mao, Phys. Rev. Lett. **94**, 185502 (2005).

[55]F. Simon and G. Glatzel, Z. Anorg. Allg. Chem **178**, 309 (1929).

[56]P. A. Urtiew and R. Grover, J. Appl. Phys. **48**, 1122 (1977).

[57]T. Sun, J. P. Brodholt, Y. G. Li, and L. Vočadlo, Phys. Rev. B **98**, 224301 (2018).

[58]E. A. Mastny and J. J. de Pablo, J. Chem. Phys. **127**, 104504 (2007).

[59]Q. J. Hong and A. van de Walle, J. Chem. Phys. **139**, 094114 (2013).

[60]Q. J. Hong and A. van de Walle, Calphad **52**, 88 (2016).